\documentclass[aps,pre,twocolumn,showpacs]{revtex4}  
\usepackage{graphicx}  
\usepackage{dcolumn}   
\usepackage{bm}        
\usepackage{amssymb}   
\usepackage{float}
\begin{document}
\title{Feedback as a mechanism for the resurrection of oscillations from death state} 
\author{V.K. Chandrasekar$^{1}$, S. Karthiga$^{2}$ and  M. Lakshmanan$^{2}$}
\address{$^1$Centre for Nonlinear Science \& Engineering, School of Electrical \& Electronics Engineering, SASTRA University, Thanjavur -613  401,Tamil Nadu, India.\\
$^2$ Centre for Nonlinear Dynamics, School of Physics, Bharathidasan University, Tiruchirappalli - 620 024, Tamil Nadu, India.}
\begin{abstract}
\par The quenching of oscillations in interacting systems leads to several unwanted situations, which necessitate a suitable remedy to overcome the quenching. In this connection, this work addresses a mechanism that can resurrect oscillations in a typical situation. Through both numerical and analytical studies, we show the candidate which is capable of resurrecting oscillations is nothing but the feedback, the one which is profoundly used in dynamical control and in bio-therapies.  Even in the case of a rather general system, we demonstrate analytically the applicability of the technique over one of the oscillation quenched states called amplitude death state.  We also discuss some of the features of this mechanism such as adaptability of the technique with the feedback of only a few of the oscillators.  
\end{abstract}
\pacs{05.45 Xt, 87.10 -e, 87.19.Ir}
\maketitle
 \section{Introduction}
\par The interaction among oscillators in a system not only leads them into a cooperative dynamics but also often quenches their oscillations.  There are two dynamically different oscillation quenching phenomena which are termed as amplitude death (AD) and oscillation death (OD) \cite{r1, r2}.  In AD, the amplitude of oscillation quenches to zero, whereas OD is caused by quenching in the frequency of oscillation \cite{r1}.  It is also defined that AD occurs via the stabilization of a homogeneous steady state (HSS) while OD occurs via the  stabilization of an inhomogeneous steady state \cite{r2, r3, c4}.  The mechanisms underlying these two quenching phenomena have been identified recently and the results show that the parametric mismatch \cite{r3}, dynamic \cite{r8,r9}, time delay \cite{r10,r11,r12} and nonlinear couplings \cite{r13} underly AD,  while OD occurs mainly because of the symmetry breaking coupling in the system \cite{r2}.   Recently, diverse routes of transition from the AD to OD have also been reported \cite{r3, r14,r15}. 
\par Experimental and theoretical studies show definitive evidence of oscillation quenching in realistic systems ranging from biological \cite{r16,r16a}, chemical \cite{r17,r17a}, electronic \cite{r18} and laser \cite{r19} systems to climate \cite{r20} systems.   Such oscillation quenching in many cases leads to undesirable situations.   In the interaction between neuronal dynamics and brain metabolism, the decrease in  cerebral metabolic rate, coupled with the stabilizing properties of ATP-gated potassium channels, leads to a burst suppression in the EEG pattern which symbolizes inactivated brain \cite{bur}.  This type of suppression results in hypothermia, coma and Ohtahara syndrome, a type of early infantile encephalopathy and is also observed during deep levels of anesthesia.  The suppression of normal sinus rhythm of pacemaker cells causes cardiac arrest \cite{r23}. 
\par Owing to the fatal consequences due to oscillation suppression, interesting efforts have been undertaken to retrieve/resurrect oscillations of the system \cite{r25,r24}.  In \cite{r25}, the oscillation death in diffusively coupled oscillators has been found to be eliminated through a spatial disorder in the form of parametric mismatch and in Ref. \cite{r24} processing delay is used to revoke oscillations successfully in delay coupled systems. 
\par Regarding the above mentioned issues, in this article, we demonstrate that the problem can also be well resolved by providing a suitable feedback in the system.  The latter can be found to be present in most of the natural systems, including neural networks \cite{a1}, genetic networks \cite{a2}, vision systems \cite{a3}, etc.   The vital role of feedback in controlling the dynamics of the given system and the control over synchronization \cite{con1, con2, con3, syn2} are already known, which can be seen in a variety of fields ranging from electronics \cite{ele}, biology to quantum information \cite{qu1,qu2}.  For example, the feedback control of deep brain simulation has been found to be the most effective treatment for chronic neural diseases like essential tremor, dystonia and Parkinson's disease \cite{dbs1,dbs2}.  Also, the feedback generated by the voltage-gated ion channels in neural cells is found to be crucial in generating neural signals \cite{bok}. 
\par  In this article, we show the applicability of the feedback technique in resurrecting oscillations in a wide range of systems.  We show both numerically and analytically that the addition of feedback destabilizes the stable attractors which results in a wiping out of the oscillation quenching and inducing a resurrection of oscillations.  In addition, by considering a rather general system, we prove analytically the above destabilizing nature over the AD state.
\par Further, it will be more important to develop an adaptable mechanism thereby improving the ones available in the literature at present.  This is because,  for example to use the available parametric mismatch method, one needs to tune the internal parameters of the system, while the processing delay also depends on the underlying process of the system where that process may be unknown in many situations so as to hinder the efficiency.  In contrast, the feedback method suggested here can be given more easily which is already in practice under different contexts such as in deep brain simulation \cite{dbs1,dbs2}.
\par  In addition to the above adaptable nature of feedback, with the aid of numerical and analytical studies, we show the important fact that this method does not impose a restriction that the output of all the oscillators need to be fedback.  From the output of only a few of the system oscillators, we show that in typical systems the resurrection of oscillations can be achieved easily. 
\par The structure of the paper is as follows.  In section \ref{genr} we present the general form of the system that we consider.  In section \ref{diffu}, we illustrate the role of feedback on two important oscillation quenching scenarios, namely the symmetry breaking coupling and the parametric mismatch in a system of diffusively coupled Stuart-Landau oscillators, through numerical analysis.  In section \ref{anal}, we present suitable analytical support of the numerical results based on an appropriate linear stability analysis.  In section \ref{dynam}, we illustrate the role of the considered feedback in indirectly coupled or dynamically coupled Stuart-Landau oscillators. A realistic chemical oscillator model, namely the Brusselator model, is considered in section \ref{brusec}.  We have also proved the applicability of the technique in more general situations over amplitude death state in Appendix \ref{gen_an}.  In addition, in the Appendix  \ref{more} we illustrate our method with different coupling schemes and with different models such as van der Pol oscillator, R\"{o}sseler system and so on.  Appendix \ref{app2_an} includes the details in obtaining the boundary curves of the AD region which were given in Sec. \ref{anal}. A summary of our results and conclusions are presented in section \ref{conclu}.
\section{\label{genr}The General model}
\par Consider a system of coupled dynamical systems, 
\begin{eqnarray}
\dot{{\bf w}_i}={\bf f}_i({\bf{w}}_i)+\epsilon \sum_{j=1}^{N}{{\bf L}_{ij}}{\bf H}({\bf{w}}_j)+ \eta {\bf g}({{\bf u}(t)}), \nonumber\\\;\;i=1,2,\ldots N
\label{gen}
\end{eqnarray}
where $ {\bf f}_i : {\bf R}^d \rightarrow {\bf R}^d$ characterizes the dynamics of the isolated $i$-th system,  ${\bf{w}}_i \in {\bf R}^d$ is a $d$ dimensional state vector of the system $i$, $ {{\bf L}_{ij}}$ is the $d\times d$ coupling matrix of the network,  $(\epsilon, \eta) \in {\bf R}$ are respectively the uniform coupling and feedback strengths, ${\bf H}: {\bf R}^d \rightarrow {\bf R}^d$ is a coupling function and $ {\bf g}({{\bf u}(t)}): {\bf R}^d \rightarrow {\bf R}^d$ is the feedback term which can be written as ${\bf g}({{\bf u}(t)})={\bf Q}{\bf u}(t)$. Here $\bf Q$ is simply a $d \times d$ constant matrix and ${\bf u}(t) \in {\bf R}^d$ characterizes the feedback and it depends on the state vectors of the system.  Such a dependence of ${\bf u}(t)$ on the state vectors of the system may be linear (Example: ${{\bf u}(t)}= \sum_{k=1}^{N} {a_k  {\bf w}_k}$) or nonlinear (Example: ${{\bf u}(t)}= \sum_{k=1}^{N} {a_k ({{\bf w}_k}^T{\bf w}_k)^q {\bf w}_k}$), where $a_k$'s represent weight factors which can take values from $0$ to $1$ and $q$ is a suitable number.  In our following study, we consider the form of ${\bf u}(t)$ as ${{\bf u}(t)}= \sum_{k=1}^{N} {a_k  {\bf w}_k}$.
\par  In the Appendix \ref{gen_an}, we have considered rather general forms for ${\bf f}_i$ and ${\bf L}_{ij}$ and shown analytically that the trivial AD state which appears in the system could be wiped out through the strengthening of $\eta$ so as to resurrect oscillations. 
\begin{figure}
\begin{center}
\hspace{-0.2cm}
   \includegraphics[width=9.6cm]{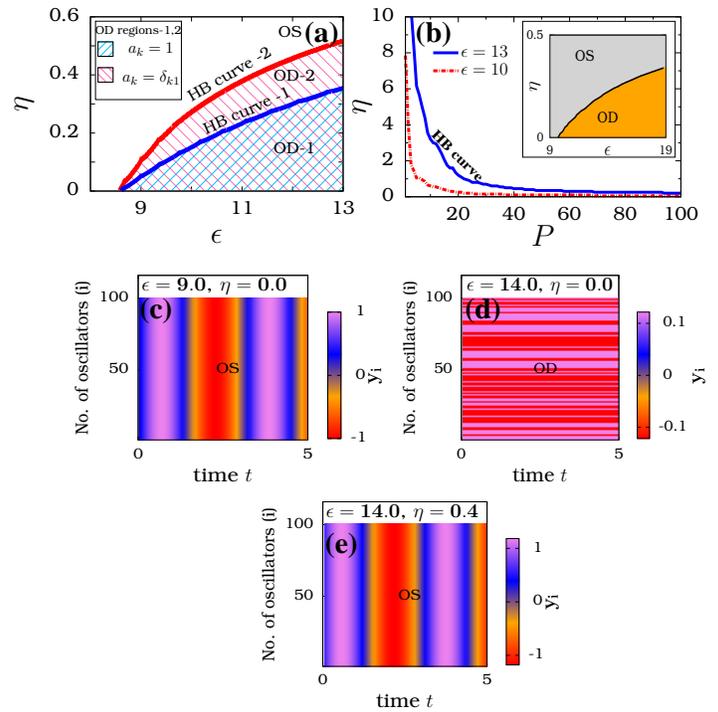}
\vspace{-0.10cm}
\end{center}
   \caption{(Color online) (a) Suppression in OD regions of the system (\ref{sys1}) with respect to the introduced two different forms of feedback (i) $a_{k}=1$ and (ii) $a_{k}=\delta_{k1}$, $k=1,2$, for the choice $N=2$ and $\omega=2.0$.  (b)For two choices of $\epsilon$, $\epsilon=10.0$ and $\epsilon=13.0$, the change in the HB point by the increment of $P$ (which appears in the expression of $a_k$) with $N=100$ in (\ref{sys1}).  Inset in (b) projects the OD and oscillatory (OS) regions in the ($\epsilon, \eta$) space for $P=N$. (c),(d) and (e): Temporal behaviors of the system in the OS, OD and revoked OS state respectively for the values of ($\epsilon, \eta$) $=$ ($9.0,0$), ($14.0,0$), and ($14.0,0.4$).  The color bar in Figs. (c), (d) and (e) gives information on the value of $y_i$ corresponding to different colors.} 
\label{am1}   
\end{figure}
\section{\label{diffu}Diffusively coupled system: Numerical Analysis} 
\subsection{\label{syyb}Symmetry breaking coupling}
\par To start with, we use the paradigmatic model known as the coupled Stuart-Landau oscillators for the purpose of illustration for the advocated feedback method. It is well known that the dynamical equation defining the Stuart-Landau oscillator can be obtained from a general ordinary differential equation near a Hopf bifurcation point \cite{rev4, new1}. As the Hopf bifurcation arises widely in the literature, the Stuart-Landau oscillator helps to model a variety of systems in different areas ranging from biology \cite{rev1, rev2,new7} to lasers \cite{lasr2, lasr} and is also used in the reaction-diffusion process \cite{rev4, rev5}.  This model is often used in neural networks to model spiking neurons \cite{rev1, rev2, new7}.  The first reason for using the model in neural networks is that the periodically spiking neurons have an exponentially stable limit cycle attractor and secondly the real part of the complex amplitude of the Stuart-Landau oscillator can describe the membrane voltage in the neurons and the imaginary part can be related to the recovery variable embedding the effects of the other variables of physiological neuron models \cite{nn1}.  Thus, in the literature we can find the use of this model in studying the effects of synchronization and desynchronization in neural networks \cite{new1, new7, new2, new5, new6, new4} and also various collective dynamical states, including chimeras \cite{chim, chim1, chim3}. In our study, we also include other useful models such as the van der Pol oscillator, R\"{o}ssler system and Brusselator model.  The corresponding results are briefly indicated in the Appendix B and Sec \ref{brusec}. 
  
\par Now, we first consider a system of coupled Stuart-Landau oscillators which is characterized by 

\begin{small}
\begin{eqnarray}
&{\bf f}_i({\bf{w}}_i)=\left(\begin{array}{cc}
  x_i-\omega_i y_i-r_i^2x_i  \\
  y_i+\omega_i x_i-r_i^2 y_i  \\
\end{array}\right),&\; \;\;{\bf H}({\bf{w_j}})=\bf{w_j},\qquad \qquad  \nonumber \\
&{{\bf L}_{ij}}=\frac{1}{N}\left(\begin{array}{cccc}
-N\delta_{ij}+1 &0  \\
 0&0 \\
\end{array}\right),&\;\;\; {\bf u}(t)= \sum_{k=1}^{N} {a_k  {\bf w}_k} \nonumber \\
&{\bf g}({\bf u}(t))={\bf Q}{\bf u}(t),&\;\; {\bf Q}=\frac{\bf I}{N},
\label{sys1}
\end{eqnarray}
\end{small}
where $r_i^2=x_i^2+y_i^2$, ${\bf{w}_i}=[x_i \; y_i]^{T}$, $\delta_{ij}$ is the Kronecker delta ($\delta_{ij}$ $=0$, if $i \neq j$ and $\delta_{ij}=1$, if $i=j$) and {\bf I} represents the identity matrix, in Eq. (\ref{sys1}).  Here the diffusive coupling acts only on the first half of the evolution equations ($x$- variable alone) which breaks the rotational symmetry \cite{r25} and consequently induces oscillation death in the system. 
\par  To elucidate clearly the role of feedback in (\ref{sys1}), we first consider the case $N=2$, with $\omega_1=\omega_2=\omega$.  In the case $\epsilon=0$, $\eta=0$, we note that the individual systems in (\ref{sys1}) show limit cycle oscillations with $|r_i|=1$.  The introduction and strengthening of diffusive coupling ($\epsilon \neq 0$) stabilizes the symmetric pair of nontrivial equilibrium points $(x^*_1,y^*_1, x^*_2, y^*_2)=$ $(a_i^*,b_i^*,-a_i^*, -b_i^*)$, where $i=1,2$, $a_{1,2}^*= c b_{1,2}^*$, $b_{1,2}^*= \pm \sqrt{\frac{1+\omega c}{1+c^2}}$ and $c=\frac{-\epsilon + \sqrt{(\epsilon^2-4\omega^2)}}{2 \omega}$, through a sub-critical Hopf bifurcation \cite{r3} which gives rise to OD.  With the introduction of feedback  $\eta \neq 0$, these nontrivial equilibrium points soon lose their stability via a super-critical Hopf bifurcation.  In our study, we introduced such a feedback in two ways: (i) $a_k=1$ and (ii) $a_k=\delta_{k1}$, $k=1,2$ (again here $\delta$ denotes the Kronecker delta).  First by setting $a_k=1$,  for different values of $\epsilon$ we traced the Hopf bifurcation points and these points are collectively shown as the HB curve$-1$ in Fig. \ref{am1}(a).  The region lying under this curve is an OD region (denoted by OD$-1$) and the parametric region above this curve is free from OD and corresponds to oscillatory states (OS).  Similarly for $a_k=\delta_{k1}$, $k=1,2$, the OD region (OD-$2$ which includes OD-$1$ also) and the curve of Hopf bifurcation points (HB curve$-2$) have been shown in fig.\ref{am1}(a), which show that the uniform distribution of $a_k$ as in (i) helps to redeem from the OD state sooner than in the case (ii).  

\par By extending the constituents of the network to $N=100$, we have verified that this technique can work as well with larger $N$.  Further, we have checked whether the feedback needs contributions from all the constituents of the network.  This is vital as in a practical situation we cannot assure or impose all the constituents to contribute to the feedback.   Thus, we have distributed $a_k$'s as  $a_k=\sum_{j=1}^{P} \delta_{kj}$, $k=1,2,...,N$, where $P$ determines the number of contributing components of the network.  By varying $P$, we have drawn the HB curves separating the OD state with the OS state for two different values of $\epsilon$, $\epsilon=10.0$ and $\epsilon=13.0$  in Fig. \ref{am1}(b).  Interestingly, these curves demonstrate clearly that the contribution from even a single oscillator is sufficient to revoke oscillations in the network.  Secondly, the critical value of $\eta$ (HB point) above which the OS state arises gets decreased sharply with that of $P$.  These facts prove that this technique can work well regardless of the number of oscillators present in the network and the number of them which contributes towards the feedback.  For simplicity, we chose $P=N$ in the following studies.  
\begin{figure}
\begin{center}
   \includegraphics[width=9.6cm]{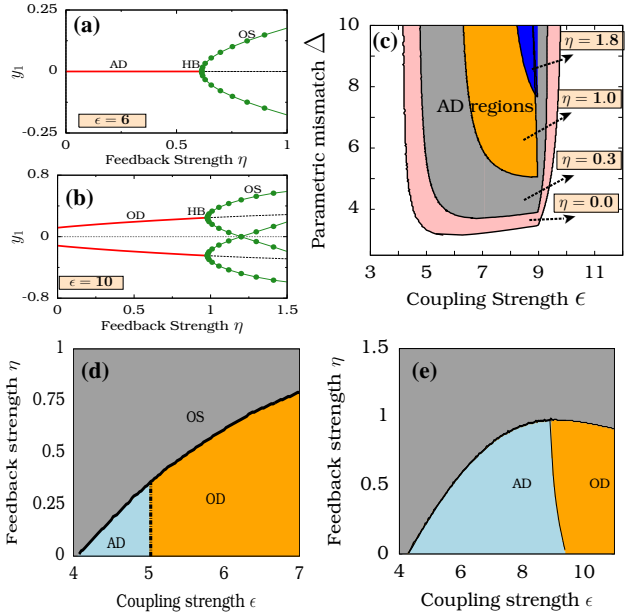}
\vspace{-0.2cm}
\end{center}
   \caption{(Color online) The transition from AD (a) or OD (b) state to OS state via Hopf bifurcation with respect to $\eta$ in the system (\ref{sys1}) endowed with a parametric mismatch of $\Delta=5.0$, $\omega_2=2.0$ and $N=2$. (c) The reduction of AD regions in the ($\Delta, \epsilon$) space with the increase of $\eta$ and by keeping $\omega_2=2$ and $N=2$. (d) and (e) The emergence of OS state from AD and OD states in $N=100$ case for $\Delta=5.0$ and $\omega_2=2.0$ which is obtained for two different set of initial conditions. AD region in Fig. (e) is also in conformity with the analytical results presented in the Sec. \ref{anal}.}
\label{bif1} 
\end{figure}
\par The inset of Fig. \ref{am1}(b) depicts the information about OD and OS regions with $N=100$ oscillators.  Correspondingly, in Figs. \ref{am1}(c), \ref{am1}(d) and \ref{am1}(e), we have captured the temporal behaviors of the system (in the $y$-variables) for different sets of ($\epsilon,\eta$), for a finite time interval after leaving out sufficiently large transients.   The first one (Fig. \ref{am1}(c)) shows the temporal behavior at an OS state of the system when $\eta=0$ and $\epsilon=9.0$, where the value of $\epsilon$ is not sufficient to induce OD.   Now increasing $\epsilon$ to $14.0$ (while keeping $\eta=0$), the subsequent figure (Fig. \ref{am1}(d)) shows the quenching of this oscillation.  Now switching $\eta$ on, the temporal behavior in Fig. \ref{am1}(e) shows the resurrected oscillations for $\eta=0.4$.
\subsection{\label{parame}Effect of parametric mismatch} 
\par Rubchinsky and Sushchik  \cite{r25} have introduced a disorder in the form of parametric mismatch which revokes oscillations  in (\ref{sys1}).  On the other hand, just like symmetry breaking which is predominant in inducing OD, the parametric mismatch is also a key candidate that induces AD in the system. Recently, Koseska {\it et al.} in \cite{r3, r4,r5} have shown that an increase in this inhomogeneity not only induces AD but also OD, whereas the feedback mechanism that we consider here does not show such a behavior in the system.  This feature provides a definitive advantage over parametric mismatch.
\par Now, we augment the system with a parametric mismatch.  For the mismatch in the parameter $\omega$, $\frac{\omega_1}{\omega_2}=\Delta=5$, in the above coupled Stuart-Landau oscillators (\ref{gen})-(\ref{sys1}) with $N=2$, we find the existence of both AD ($\epsilon=6.0$) and OD ($\epsilon=10.0$) while $\eta=0$.  Now, from these death states the transitions towards OS state by $\eta$ are demonstrated in  Figs. \ref{bif1}(a) and \ref{bif1}(b), which show the destabilization of both the AD and OD states via supercritical Hopf bifurcations.  Further, the role of $\eta$ over $\Delta$ and $\epsilon$ is more clearly demonstrated in Fig. \ref{bif1}(c), where the colored islands denote the AD regions for the values of $\eta$ $= 0.0,0.3,1.0$ and $1.8$.  One can check that similar phenomenon occurs to OD regions also which we do not depict here explicitly.
\par  Now considering the case of $N=100$ globally coupled oscillators with $\omega_i=10.0$ for $i=1,2,...,50$ and $\omega_i=2.0$ for $i=51,52,...,100$ in (\ref{sys1}), we have demonstrated the reduction in the AD and OD regions in Figs. \ref{bif1}(d) and \ref{bif1}(e).  The above two figures are plotted for two different sets of initial conditions.  Among them, the AD region in Fig. \ref{bif1}(e) is the analytically relevant region which has been obtained in the next section.
\par In the next section, the above obtained numerical results on the system (\ref{sys1}) are verified through analytical results wherever possible.  Also, we illustrate the applicability of the technique to several situations in Appendix \ref{more}, where we considered (i) Repulsive link: Stuart-Landau (SL) oscillator, (ii) Conjugate coupling: SL oscillator, (iii) Repulsive link: van der Pol oscillator, (iv) Directly and indirectly coupled R\"ossler system and (v) Other chaotic oscillators (Sprott and Lorenz systems).

\section{\label{anal} Analytical confirmation of suppression of death states}
\begin{figure}
\hspace{-1.0cm}
\begin{center}
   \includegraphics[width=9.7cm]{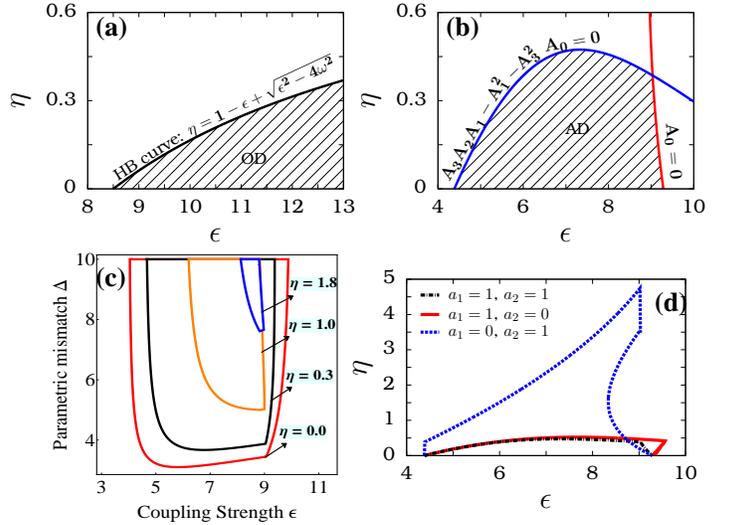}
\end{center}
   \caption{(Color online) Without parametric mismatch case: Fig. (a) Analytical results of the OD regions in ($\epsilon,\eta$) space for $N=2$ oscillators in system (\ref{sys1}) with $\omega_1=\omega_2=2.0$. With parametric mismatch case: Fig. (b) AD regions of system (\ref{sys1}) for $\Delta=4$, $\omega_2=2$ and $a_1=a_2=1$ bounded by the curves defined in (\ref{condi}). Fig. (c) Analytically obtained boundaries of AD regions in ($\Delta$, $\epsilon$) for different values of $\eta$ (=$0,0.3,1.0,1.8$), where $a_1=a_2=1$. Fig. (d) HB curves for the three different cases (i) $a_1=a_2=1$, (ii) $a_1=1,\, a_2=0$ and (iii) $a_1=0, \, a_2=1$ for $\Delta=4$ and $\omega_2=2.0$.}
\label{anal1} 
\end{figure}
\par In this section, we present relevant analytical confirmations of the numerical results corresponding to the system (\ref{sys1}) for the cases with and without parametric mismatch discussed earlier in Sec. \ref{diffu}.   In addition through the obtained analytical results, we show the effectiveness of the technique with the feedback contribution coming from a few number of oscillators in the network.
\par \subsection{Without parametric mismatch:  $N=2$ case}
\par In the absence of any parametric mismatch, as pointed out earlier the system in (\ref{sys1}) has a trivial equilibrium point $E_0$: $(0,0,0,0)$ and two pairs of  non-trivial equilibrium points as
\begin{eqnarray}
 &&E_i: (x^*_1,y^*_1, x^*_2, y^*_2)= (a_i^*,b_i^*,-a_i^*, -b_i^*), \quad i=1,2,3,4, \nonumber \\
&&\mathrm{where} \nonumber \\
&&a_{1,2}^*= c b_{1,2}^*,\; b_{1,2}^*= \pm \sqrt{\frac{1+\omega c}{1+c^2}}, \nonumber \\
&& \;a_{3,4}^*= d b_{3,4}^*, b_{3,4}^*= \pm \sqrt{\frac{1+\omega d}{1+d^2}},
\label{eqp}
\end{eqnarray}
 where $c=\frac{-\epsilon + \sqrt{(\epsilon^2-4\omega^2)}}{2 \omega}$ and $d=\frac{-\epsilon - \sqrt{(\epsilon^2-4\omega^2)}}{2 \omega}$.  The  linear stability of these fixed points is determined by the eigenvalues of the Jacobian matrix
\begin{small}
\begin{eqnarray}
J=\left(\begin{array}{cccc}
A_1 & -\omega-2 x_1^* y_1^*& \frac{\epsilon}{2}+\frac{\eta}{2}&0 \\
\omega-2 x_1^* y_1^*&B_1& 0 &\frac{\eta}{2} \\
\frac{\epsilon}{2}+\frac{\eta}{2}& 0 &A_2& -\omega - 2 x_2^* y_2^* \\
0&\frac{\eta}{2} & \omega-2 x_2^* y_2^* & B_2
\end{array}\right)
\end{eqnarray}
\end{small}
where $A_i=1-3{x_i^*}^2-{y^*_i}^2-\frac{\epsilon}{2}+\frac{\eta}{2}$, $B_i=1-{x_i^*}^2-3 {y_i^*}^2+\frac{\eta}{2}$, $i =1,2$.  While $\eta=0$, the eigenvalues of $J$ corresponding to the trivial equilibrium point $E_0$ are 
\begin{eqnarray}
\mu_{1,2}^{(0)}= \frac{1}{2}(2-\epsilon \pm \sqrt{\epsilon^2 -4 \omega^2}), \quad \mu_{3,4}^{(0)}=1\pm i \omega.
\label{0eig}
\end{eqnarray}
The eigenvalues corresponding to the equilibrium points $E_{1}$ and $E_{2}$ are given by
\begin{small}
\begin{eqnarray}
\mu_{1}^{(j)}&=&-\sqrt{\epsilon^2-4 \omega^2}, \quad \mu_2^{(j)}= (\epsilon -2 )-\sqrt{\epsilon^2 - 4 \omega^2}, \nonumber \\
\mu_{3,4}^{(j)}&=&\left((\epsilon-1)-\sqrt{\epsilon^2 -4 \omega^2}\right) \pm \sqrt{C_1} \quad , \;\; j=1,2 \label{1eig} \\
\mathrm{where}&& \hspace{5cm} \nonumber \\
C_1&=&\frac{-2(\epsilon-1)+\sqrt{\epsilon^2-4 \omega^2}(\sqrt{\epsilon^2-4 \omega^2}-(\epsilon-2))}{2}. \label{c11}
\end{eqnarray}
\end{small}
 Similarly, $E_{3}$ and $E_{4}$ have the set of eigenvalues
\begin{small}
\begin{eqnarray}
\mu_{1}^{(j)}&=&\sqrt{\epsilon^2-4 \omega^2}, \quad \mu_2^{(j)}= (\epsilon -2 )+\sqrt{\epsilon^2 - 4 \omega^2}, \nonumber \\
\mu_{3,4}^{(j)}&=&\left((\epsilon-1)+\sqrt{\epsilon^2 -4 \omega^2}\right) \pm \sqrt{C_2}. \;\; j=3,4, \\
\mathrm{where}&& \hspace{5cm} \nonumber \\
C_2&=&\frac{-2(\epsilon-1)+\sqrt{\epsilon^2-4 \omega^2}(\sqrt{\epsilon^2-4 \omega^2}+(\epsilon-2))}{2}. 
\label{2eig}
\end{eqnarray}
\end{small}
 From Eqs. (\ref{0eig}-\ref{2eig}) we can note that among the five equilibrium points, $E_{1}$ and $E_{2}$ are found to have all their eigenvalues satisfying the condition $Re[\mu]<0$ for the parametric range $\epsilon > \frac{(1+4 \omega^2)}{2}$ while $\omega> \sqrt{\frac{1}{3}}$, $\frac{1+4 \omega^2}{2}<\epsilon<\frac{2}{3}(2-\frac{2}{3}\sqrt{1-3 \omega^2})$ while $\frac{1}{2} < \omega <\sqrt{\frac{1}{3}}$ and $\epsilon>\frac{2}{3}(2+\frac{2}{3}\sqrt{1-3 \omega^2})$ while $\omega<\sqrt{\frac{1}{3}}$ and thus they are stable in this range.  On the other hand the other equilibrium points $E_0$, $E_3$ and $E_4$ can never become stable for any choice of parametric values.  Thus the stabilization of $E_1$ and $E_2$ essentially gives rise to oscillation death in the system. 
\par Now we introduce the feedback in such a way that $a_1$ and $a_2$ of ${\bf u}(t)$ in (\ref{sys1}) take the values $a_1=1$ and $a_2=1$.  With such a choice, we find that the stability determining eigenvalues corresponding to the equilibrium points change as  
\begin{eqnarray}
\widetilde{\mu}_{1,2}^{(j)}=\mu_{1,2}^{(j)}, \;\; \mathrm{and} \;\;\widetilde{\mu}_{3,4}^{(j)}=\mu_{3,4}^{(j)}+\eta, \;\; j=0,1,2,3,4.
\label{eigg}
\end{eqnarray}
The above equation shows that an increase in $\eta$ can destabilize the equilibrium points $E_1$ and $E_2$ through a Hopf bifurcation.  In the cases where $\omega \geq 1$, for all values of $\epsilon$ the Hopf bifurcation occurs at
\begin{eqnarray}
 \eta=(1-\epsilon)+\sqrt{\epsilon^2 -4 \omega^2}, 
\label{etac1}
\end{eqnarray}
whereas in the case of $\omega <1$, if $\epsilon>\frac{(2 \omega^2 -2 \omega+1)}{1-\omega}$, the Hopf bifurcation occurs at
\begin{eqnarray}
\eta=(1-\epsilon)+\sqrt{\epsilon^2 -4 \omega^2}-\sqrt{C_1}
\label{etac2}
\end{eqnarray} 
where $C_1$ is given in (\ref{c11}). If $\epsilon<\frac{(2 \omega^2 -2 \omega+1)}{1-\omega}$, the Hopf bifurcation occurs as in (\ref{etac1}).
 The other equilibrium points $E_0$, $E_3$ and $E_4$ are found to remain unstable. For the case of $\omega=2.0$, the curve of Hopf bifurcation points ($\eta=1-\epsilon+\sqrt{\epsilon^2 -4 \omega^2}$) separating the death regimes with the oscillatory regimes is shown in Fig. \ref{anal1}(a) which matches exactly with the one obtained numerically (HB curve - $1$ in Fig. \ref{am1}(a)).  The analytical treatment of the other case corresponding to the HB curve-2 can also be investigated in a similar manner, though the results cannot be written down in such a transparent  manner.  So we do not present the details here.   
\subsection{\label{par_ann2} With parametric mismatch: $N=2$ case}
\par Next with the introduction of a parametric mismatch in the system (\ref{sys1}), we prove the validity of the technique for the trivial AD state analytically for the $N=2$ case.  As mentioned earlier in Sec. \ref{parame}, an increase in the value of the parametric mismatch parameter $\Delta$ causes the stabilization of the trivial equilibrium point $(0,0,0,0)$ and thus introduces AD in the system. To destabilize the latter, we first introduce the feedback in such a way that $g({\bf u}(t))=\frac{1}{2}{\bf u}(t)$, where ${\bf u}(t)=a_1 {\bf w}_1+a_2 \bf{w}_2$. 
\par As before, through a linear stability analysis of the system (\ref{sys1}) with parametric mismatch included, we look for the stable regions of the equilibrium point $(0,0,0,0)$, which can be studied from the characteristic equation of the linear eigenvalue problem of the system as
\begin{eqnarray}
 \mu^4+A_3 \mu^3+A_2 \mu^2+A_1 \mu+A_0=0.
\label{eveq}
\end{eqnarray}  
\begin{figure}
\begin{center}
\hspace{-01.2cm}
   \includegraphics[width=9.6cm]{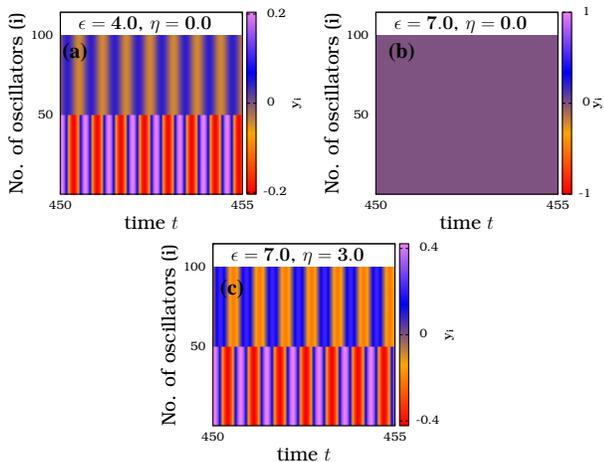}
\vspace{-0.10cm}
\end{center}
   \caption{(Color online) Temporal behaviors of the system (\ref{sys1}) with $N=100$ oscillators in the OS, OD and revoked OS states respectively, for the values of ($\epsilon, \eta$) $=$ ($4.0,0.0$), ($7.0,0.0$), and ($7.0,3.0$).  Here, we have taken $N_1=50$, $N_2=50$, $p_1=20$, $p_2=25$, $\omega_2=2.0$ and $\omega_1=10.0$.   The color bars in Figs. (a), (b) and (c) represent the values of the variables $y_i$.}
\label{sync}   
\end{figure} 
The coefficients  $A_3$, $A_2$, $A_1$ and $A_0$ in the above equation are given by
\begin{small}
\begin{eqnarray}
A_3&=&\epsilon-4-\tilde{a}\eta, \nonumber \\
A_2&=& \frac{1}{4} \tilde{a}^2 \eta^2+\tilde{a} \eta (3 - \epsilon)+(6-3 \epsilon+\omega_1^2+\omega_2^2), \nonumber \\
A_1&=&A_{11} \eta^2+A_{12} \eta +A_{13}, \qquad \nonumber \\
A_0&=&A_{01} \eta^2+ A_{02} \eta+A_{03}, \qquad  \nonumber \\
\mathrm{where}&&\hspace{5cm} \nonumber \\
A_{11}&=&\frac{1}{4} (\epsilon-2) \tilde{a}^2,\nonumber \\
A_{12}&=& (2 \epsilon - 3) \tilde{a}-(\omega_1^2 a_2+\omega_2^2 a_1), \nonumber \\
A_{13}&=& (3 \epsilon-4)+\frac{1}{2} (\epsilon-4)(\omega_1^2+\omega_2^2), \nonumber \\
A_{01}&=&\frac{1}{4} \left[(1-\epsilon) \tilde{a}^2+(\omega_1 a_2+\omega_2 a_1)^2 \right], \nonumber \\
A_{02}&=&\frac{1}{4}\left[(4(1-\epsilon)+\epsilon \omega_1 \omega_2) \tilde{a}-(\epsilon-4)(\omega_1^2 a_2+\omega_2^2 a_1) \right], \nonumber \\
A_{03}&=&(1+\omega_1^2)(1+\omega_2^2)-\frac{\epsilon}{2} (2+\omega_1^2+\omega_2^2).
\label{coeff}
\end{eqnarray}
\end{small}
Here $\tilde{a}=a_1+a_2$.  Since the characteristic equation for the eigenvalues is quartic in nature, we use the well known Routh-Hurwitz criteria \cite{sir_bok} to obtain the stable AD regions of the system.  By doing so, we find that the AD regions are bounded by the curves 
\begin{eqnarray}
&&A_0=0 \quad \mathrm{or} \quad \eta=\frac{-A_{02} \pm \sqrt{A_{02}^2-4 A_{01} A_{03}}}{2  A_{01}} \nonumber \\
\mathrm{and}&& \hspace{5cm} \nonumber \\
&&A_3 A_2 A_1- A_1^2-A_3^2 A_0=0.
\label{condi}
\end{eqnarray}
The details of obtaining the boundary curves from the R-H criteria are presented in the Appendix \ref{app2_an}, and the AD region bounded by the curves (\ref{condi}) has been shown in Fig. \ref{anal1}(b) for $a_1=1$ and $a_2=1$.  Fig. \ref{anal1}(c) which portrays the boundaries of the AD regions in the ($\Delta, \epsilon$) space clearly shows that the analytical results match nicely with that of the numerical results given in Fig. \ref{bif1}(c). 
\par Next by varying the nature of oscillators contributing towards feedback we have plotted Fig. \ref{anal1}(d), where we considered three cases (i) $a_1=1$, $a_2=1$ (both the oscillators contributing) (ii) $a_1=1$, $a_2=0$ (only the high frequency oscillator contributing) and (iii) $a_1=0$, $a_2=1$ (only the low frequency oscillator contributing).  The analytically obtained Hopf bifurcation curves for all the above three cases have been presented in Fig. \ref{anal1}(d).  From the figure, we can note that for the first two cases (i) and (ii) the OS state gets revoked from the AD state for even small values of $\eta$ and the Hopf bifurcation curves of these two cases are closer to each other.   But, in the case where the low frequency oscillator alone is contributing, we find that comparatively higher values of $\eta$ are needed to revoke oscillations.  This shows that for a quicker resurrection of oscillations, the feedback from the high frequency oscillator is preferable.  However, the resurrection of oscillations is possible even if one of the oscillators is contributing towards the feedback.

\subsection{\label{par_ann100}With parametric mismatch: $N=100$ case}
Next, we extend our studies on the revocation of oscillations from the AD state for the case of $N=100$ oscillators, where the parametric mismatch in the system is introduced in such a way that the system has two groups of oscillators.  The first group contains $N_1$ oscillators with $\omega_i=\omega_1$, $i=1,2,...,N_1$ and the second group has $N_2$ oscillators with $\omega_i=\omega_2$, $i=1,2,...,N_2$ with $\omega_1> \omega_2$ and $N_1+N_2=N$.  Also, we consider that among the $N_1$ oscillators in the first group only the output of a sub-group of $p_1$ oscillators is fed back, in other words, the $a_k$'s of ${\bf u}(t)$ in (\ref{sys1}) take the values as
\begin{eqnarray}
 a_k&=&1, \;\; \mathrm{for} \;\; k=1,2,...,p_1, \nonumber \\
 a_k&=&0, \;\; \mathrm{for} \;\; k=p_1+1,p_1+2,...,N_1. 
\label{ak1}
\end{eqnarray}
 Similarly in the second group of $N_2$ oscillators only the output of $p_2$ oscillators is fedback or 
\begin{eqnarray}
a_k&=&1, \;\; \mathrm{for} \;\; k=N_1+1,N_1+2,...,N_1+p_2, \nonumber \\
a_k&=&0, \;\; \mathrm{for} \;\; k=N_1+p_2+1,N_1+p_2+2,...,N_2.
\label{ak2}
\end{eqnarray}
 The total number of oscillators contributing towards feedback is $N_p=p_1+p_2$. 
\par The results corresponding to $N_1=N_2=50$ oscillators presented earlier in Sec. \ref{parame} clearly demonstrate the appearance of AD state in the system.  The temporal behavior of the system in the original OS state, AD state and revoked OS state are shown in Fig. \ref{sync}, which shows the coherent nature among the oscillators in the first and second group. Next to study the case of the $N=100$ oscillators analytically, we first try to reduce the problem to a simpler level. 

\begin{figure}
\begin{center}
\hspace{-1cm}
   \includegraphics[width=9.5cm]{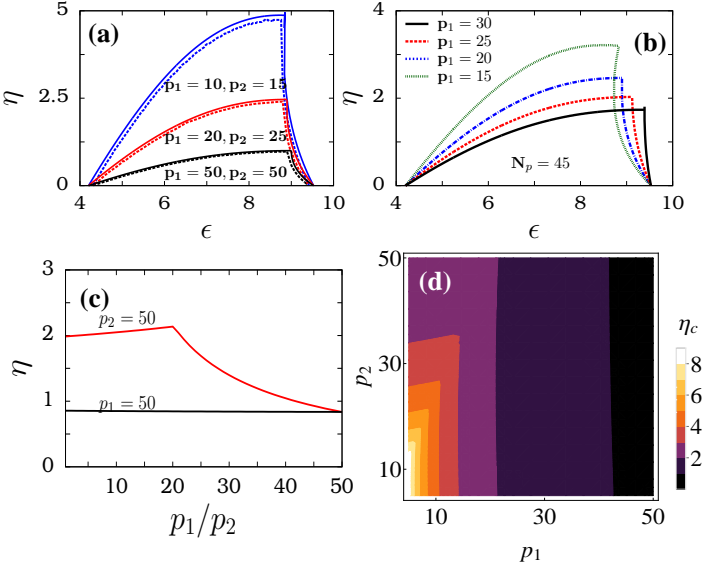}
\end{center}
   \caption{(Color online) Fig. (a) shows the numerical (continuous line) and analytical (dotted line) results on the boundaries of AD regions in ($\epsilon,\eta$) space for the values of ($p_1,p_2$)$=$ $(50,50)$, $(20,25)$ and $(10,15)$.  Fig. (b) shows the curves of $\eta_c$ for various combinations of ($p_1$, $p_2$) with fixed $N_p=p_1+p_2=45$ and $\epsilon=7.0$. Fig. (c) Fixing $p_1=50$ or $p_2=50$, the values of $\eta_c$ for different values of $p_2$ or $p_1$ is plotted with $\epsilon=7.0$.   Fig. (d) shows the value of $\eta_c$ for different values of $p_1$ and $p_2$ for $\epsilon=7$, where the values of $\eta_c$ are represented by the color function. In the above figures, we make the choice $N_1=N_2=50$, $\omega_1=10.0$ and $\omega_2=2.0$}
\label{an_ne} 
\end{figure}
\par Due to the existence of coherence among the oscillators in the first and second group, we represent the state of the oscillators in the first group by $z_1$ and the state of the oscillators in the second group by $z_2$ 
\begin{small}
\begin{eqnarray}
&{\bf f}_{i}({\bf{z}}_{i})=\left(\begin{array}{cc}
  X_{i}-\omega_{i} Y_{i}-R_{i}^2X_{i}  \\
  Y_{i}+\omega_{i} X_{i}-R_{i}^2 Y_{i}  \\
\end{array}\right),&\; \;\;{\bf H}({\bf{z_{j}}})=\bf{z_{j}},\qquad \qquad  \nonumber \\
&{{\bf L}_{ij}}=\left(\begin{array}{cccc}
(-2\delta_{ij}+1) (1-p) &0  \\
 0&0 \\
\end{array}\right),&\;\;\; {\bf u}(t)= \sum_{k=1}^{2} {a_k  {\bf z}_k} \nonumber \\
&{\bf g}({\bf u}(t))={\bf Q}{\bf u}(t),&\;\; {\bf Q}={\bf I}.
\label{sys_app}
\end{eqnarray}
\end{small}
where ${\bf{z}_i}=[X_i \; Y_i]^{T}$ and $R_i^2=X_i^2+Y_i^2$, $i=1,2$.  The coefficients in the feedback take the form $a_1=\alpha p$ and $a_2=\beta (1-p)$, where $p=\frac{N_1}{N}$, which is the ratio of oscillators present in the first group, and $1-p=\frac{N_2}{N}$ is the ratio of oscillators present in the second group.  $\alpha$ and $\beta$ are the ratio of the oscillators contributing towards feedback from the first and second groups, respectively.  Thus $\alpha=\frac{p_1}{N_1}$ and $\beta=\frac{p_2}{N_2}$, where $p_1$ and $p_2$ are respectively the number of oscillators in the first and second groups that are contributing towards feedback.  The stability of the system (\ref{sys_app}) around the equilibrium point $(0,0,0,0)$ can be studied similar to the previous case and one can obtain the characteristic equation for the eigenvalues of the above system (\ref{sys_app}) as 
\begin{eqnarray}
 \mu^4+B_3 \mu^3+B_2 \mu^2+B_1 \mu+B_0=0.
\label{eigb2}
\end{eqnarray}
The coefficients in the above equation are given by
\begin{eqnarray}
B_3&=&(\epsilon-4)-2\tilde{P} \eta, \nonumber \\
B_2&=&\tilde{P}^2 \eta^2-2( \epsilon-3)\tilde{P}\eta +(6-3\epsilon)+\omega_1^2+\omega_2^2, \nonumber \\
B_1&=&B_{11}\eta^2+B_{12}\eta+B_{13}, \nonumber \\
B_0&=&B_{01}\eta^2+B_{02}\eta+B_{03}, \nonumber \\
\mathrm{where}&&\hspace{6cm} \nonumber \\
B_{11}&=&(\epsilon-2)\tilde{P}^2, \nonumber \\
B_{12}&=&2 \beta (1-p)(-3+2 \epsilon-\omega_1^2)+2 \alpha p(-3+2 \epsilon-\omega_2^2), \nonumber \\
B_{13}&=&(3 \epsilon+2)-2(\omega_1^2 +\omega_2^2)+\epsilon(p \omega_1^2+(1-p) \omega_2^2),\nonumber \\
B_{01}&=&(1-\epsilon)\tilde{P}^2+(p \alpha \omega_2+(1-p) \beta \omega_1)^2, \nonumber \\
B_{02}&=&2\tilde{P}(1-\epsilon)+2(\beta(1-p)\omega_1^2+p\alpha \omega_2^2) \nonumber \\
&&+ \epsilon p(1-p)(\omega_1- \omega_2)(\beta \omega_1 -\alpha \omega_2), \nonumber \\
B_{03}&=&(1+\omega_1^2)(1+\omega_2^2)-\epsilon\left(1+p \omega_1^2+(1-p)\omega_2^2\right), 
\label{bs}
\end{eqnarray}
where $\tilde{P}=(p \alpha+(1-p) \beta)$.
As in the previous case, using the R-H criteria we determine the AD regions of the system.  Consequently, we find that the AD regions are  bounded by the curves defined by
\begin{eqnarray}
&&B_0=0 \quad \mathrm{or} \quad \eta=\frac{-B_{02} \pm \sqrt{B_{02}^2-4 B_{01} B_{03}}}{2  B_{01}} \nonumber \\
\mathrm{and}&&\hspace{5cm} \nonumber \\
&&B_3 B_2 B_1- B_1^2-B_3^2 B_0=0
\label{condi2}
\end{eqnarray} 
Next, using the above relations we find the AD regions in the different cases of the system.  We consider here only two cases, (i) $N_1=N_2$, (ii) $N_1 \neq N_2$.

 \par {\it{Case-1: $N_1=N_2=50$.}}~~In this case, the population of the high frequency oscillators ($N_1$) and the population of the low frequency oscillators ($N_2$) are equal (Note that $\omega_1> \omega_2$).  First, we check the consistency of the obtained analytical results with the numerical results.  For $N_1=N_2=50$, we have plotted the boundaries of the AD regions obtained from analytical and numerical studies for different values of $p_1$ and $p_2$ in Fig. \ref{an_ne}(a). (Note that for $p_1=N_1=50$ and $p_2=N_2=50$, the numerical results have been given in Fig. \ref{bif1}(e) of Sec. \ref{parame}).  The latter shows the consistency between the numerical and analytical results. 
\par Now, we look at the preferential feedback configuration for quicker resurrection of oscillations.  For the purpose, we fix the total number of oscillators contributing towards feedback as $N_p=45$ and vary the number of oscillators contributing from the first group ($p_1$) and from the second group ($p_2$).  Fig. \ref{an_ne}(b) shows the boundaries of the AD regions for different values of $p_1$ (also $p_2=N_p-p_1$), where we can observe that on increasing the contributions from the high frequency oscillators (or $p_1$), the resurrection of oscillations occur for lower values of $\eta$. Again in Fig. \ref{an_ne}(c), we first fixed $p_1=50$ and plotted the critical value of $\eta$ needed for resurrection of oscillations ($\eta_c$) for different values of $p_2$.  Similarly, we fixed $p_2=50$ and plotted $\eta_c$ for different values of $p_1$ in the same Fig. \ref{an_ne}(c).  From the figure, we find that a considerable decrease in the value of $\eta_c$ occurs only when $p_1$ is varied.  The above results show that a feedback from the high frequency oscillators is more preferable for a quicker resurrection of oscillations.  They are also evident from the Fig. \ref{an_ne}(d), where $\eta_c$ is represented by a color function.  In the figure, one can find that the rate of change in the value of $\eta_c$  is larger along the $p_1$ direction. 
  
 \par
\begin{figure}
\hspace{0.0cm}
\begin{center}
   \includegraphics[width=9.7cm]{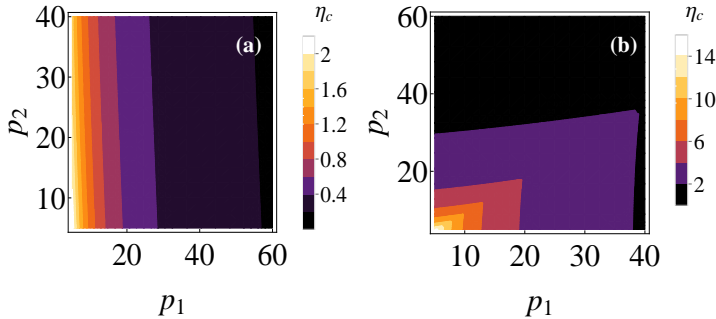}
\end{center}
   \caption{(Color online) The critical values of $\eta_c$ for different values of $(p_1,p_2)$ for the case (a) $N_1>N_2$ ($N_1=60$, $N_2=40$), (b) $N_1<N_2$ ($N_1=40$, $N_2=60$), with $\omega_1=10.0$ and $\omega_2=2.0$. }
\label{an_ne2} 
\end{figure}

 {\it{Case-2: $N_1 \neq N_2$.}}~~Just as the technique shows preference over the feedback of high frequency oscillators, it shows dependence over the population in the two groups of oscillators.  To illustrate the above, we considered two situations (i) $N_1$ $>$ $N_2$, (ii) $N_1$ $<$ $N_2$. Fig. \ref{an_ne2}(a) has been plotted for the case (i) where $N_1=60$ and $N_2=40$.  Similarly Fig. \ref{an_ne2}(b) corresponds to the case (ii) where $N_1=40$ and $N_2=60$.  The values of $\eta_c$ for different values of $p_1$ and $p_2$ are presented in Fig. \ref{an_ne2}.  Comparing Fig. \ref{an_ne2}(a) with Fig. \ref{an_ne2}(b), we find that the values of $\eta_c$ are larger in the case of $N_1< N_2$ than that in the case $N_1>N_2$.  Thus in the case where lower frequency oscillators are highly populated than the high frequency oscillators, we require stronger feedback to revoke oscillations.

\section{\label{dynam}Dynamically coupled systems} 
\par Following the studies on directly coupled systems, we turn to check the validity of the proposed scheme to indirectly coupled oscillatory systems.  For this purpose, we consider a collective system coupled to a dynamic environment, where the combined set of dynamical equations is characterized by  
\begin{figure}
\begin{center}
\hspace{-0.0cm}
   \includegraphics[width=9.6cm]{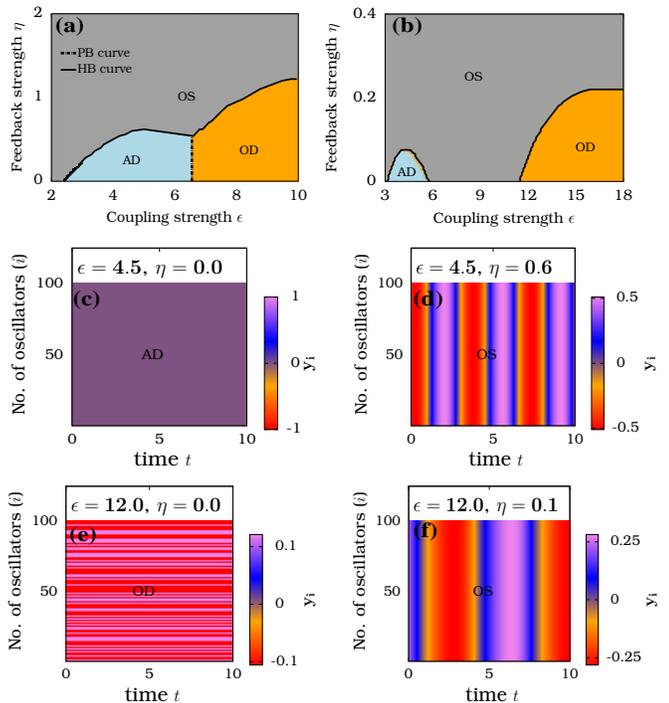}
\vspace{-0.0cm}
\end{center}
   \caption{(Color online) (a) The OD and AD regions in the ($\epsilon,\eta$) space corresponding to the system (\ref{sys2}) with $N=2$ and $\omega=3.0$, (b) The reduction in the AD and OD regions in the system with $N=100$ coupled oscillators in the ($\epsilon,\eta$) space for $\omega=3.0$. (c), (d), (e) and (f)  The corresponding temporal behavior of the system (\ref{sys2}) of $N=100$ oscillators in the AD, OD and the revoked oscillatory state.}
\label{bif2} 
\end{figure}
\begin{small}
\begin{eqnarray} 
&&{\bf f}_i({\bf{w}}_i)=\left(\begin{array}{cc}
  x_i-\omega y_i-r_i^2x_i  \\
  y_i+\omega x_i-r_i^2 y_i  \\
    -v_i\\
\end{array}\right), \qquad \qquad \nonumber \\
&&{\bf H}({\bf{w_j}})={\bf{w_j} }, \;\; \;
{{\bf L}_{ij}}=\left(\begin{array}{ccc}
   -\delta_{ij} &0 &\delta_{ij} \\
 0&0&0 \\
\frac{(1-\delta_{ij})}{N \epsilon }&0&0\\
\end{array}\right), \nonumber \\
&&{\bf g}({\bf u}(t))={\bf Q}{\bf u}(t), \;\; \;
{\bf Q}=\frac{1}{N}\left(\begin{array}{ccc}
   1 &0 &0 \\
 0&1&0 \\
0&0&0\\
\end{array}\right).
\label{sys2}
\end{eqnarray}
\end{small}

\par Here, the state of the system along with the environment is defined by the state vector ${\bf{w}}_i=\left[x_i\;\;\; y_i \;\;\; v_i \right]^T$, where the variables $x_i$ and $y_i$ correspond to the system and $v_i$ represent the environment.  While $\eta=0$,  the increase in $\epsilon$ causes a stabilization of the trivial equilibrium point ($x_i^*,y_i^*,v_i^*$)=($0,0,0$) and a further increase in $\epsilon$ stabilizes a pair of nontrivial equilibrium points defined by ($x_i^*,y_i^*,v_i^*$)=($(-1)^{i} a^*,(-1)^{i} b^*,(-1)^{i} c^*$), where $a^*$= $k b^*$, $b^*=\pm \sqrt{\frac{1+\omega k}{1+k^2}}$, $c^*=-\frac{a^*}{2}$ and $k=\frac{1}{4 \omega}(-3\epsilon + \sqrt{9 \epsilon^2 -16 \omega^2})$ via pitchfork bifurcation \cite{r14}.
\par The introduction of the feedback, $\eta \neq 0$, destabilizes both the AD and OD states via Hopf bifurcation.  Such a reduction in the territories of AD and OD states is illustrated in the ($\epsilon, \eta$) space in Fig.\ref{bif2}(a).   The curve made up of the Hopf bifurcation points separates out the AD and OD regions with the OS region. 
\par By extending $N$ to $100$, Fig. \ref{bif2}(b) depicts the AD and OD regions of the system.  When $\eta=0$, the AD state which arises by increasing $\epsilon$ is found to disappear with an increase of $\epsilon$.  But for larger $\epsilon$, we find the appearance of the OD state.  Now by switching $\eta$ on, both the AD and OD states are shown to be wiped out simultaneously.  The temporal behaviors of the system in the AD state and the resurrected oscillatory state which arise through the enhancement of $\eta$ are shown in Figs. \ref{bif2}(c) and \ref{bif2}(d).  Similarly, Figs. \ref{bif2}(e) and \ref{bif2}(f) show the behavior of the system at the OD state and the resurrection of oscillations by an increase in $\eta$.
\section{\label{brusec}Coupled Brusselator oscillators}
\begin{figure}
\begin{center}
\hspace{-0.0cm}
   \includegraphics[width=9.6cm]{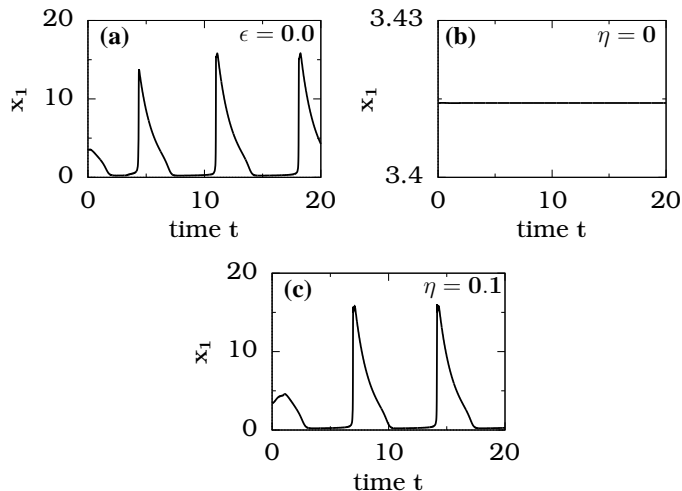}
\vspace{-0.0cm}
\end{center}
   \caption{(Color online) (a) The existence of stable limit cycle oscillations in the isolated Brusselator oscillator (\ref{bro}) for $A=2$ and $B=10$. (b) The occurrence of OD  through the coupling of the system for $\epsilon=0.5$ and $\eta=0$, and (c) The resurrection of oscillations by introducing feedback $\eta=1$.}
\label{br} 
\end{figure}
\par In this section, we consider an interesting coupled chemical oscillator modeled by the Brusselator model \cite{bru}. We consider the case where two identical cells are coupled.  In such a case, the functions in Eq. (\ref{gen}) describing the system are given by 
\begin{small}
\begin{eqnarray}
&{\bf f}_i({\bf{w}}_i)=\left(\begin{array}{cc}
 -(B+1) x_i+x_i^2 y_i +A \\
 B x_i-x_i^2 y_i\\
\end{array}\right),&\; \;\;{\bf H}({\bf{w_j}})=\bf{w_j},\qquad \qquad  \nonumber \\
&{{\bf L}_{ij}}=\left(\begin{array}{cccc}
-2\delta_{ij}+1 &0  \\
 0&-2\delta_{ij}+1\\
\end{array}\right),&\;\;\; {\bf u}(t)= \sum_{k=1}^{2} {a_k  {\bf w}_k} \nonumber \\
&{\bf g}({\bf u}(t))={\bf Q}{\bf u}(t),&\;\; {\bf Q}=\frac{\bf I}{2}.
\label{bro}
\end{eqnarray}
\end{small}
\par In the absence of the coupling, the system shows stable limit cycle oscillations which has been illustrated in Fig. \ref{br}(a).  By introducing the coupling, the system tends to an inhomogeneous steady state \cite{bru}.  For example, for $A=2$, $B=10$ and $\epsilon=0.5$, the system tends to an inhomogeneous steady state which has been illustrated in Fig. \ref{br}(b).   In such a realistic example, by introducing the feedback, we found that the oscillations are revoked by increasing $\eta$ and it has been illustrated in Fig. \ref{br}(c).

\section{\label{conclu}Conclusion}
\par From a knowledge of the role of feedback in controlling the dynamics and the coherent activities of the system such as synchronization \cite{con1, con2, con3}, we have here analyzed whether it can control oscillation quenching tendencies. For this purpose, we have demonstrated the effect of feedback over quenching induced by parametric mismatch and symmetry breaking (the key candidates for inducing AD and OD), indirect coupling and for some more cases (given in Appendix \ref{more}), through numerical as well through analytical studies wherever possible. 
 \par Further, through analytical studies on AD state of a more general system (Appendix \ref{gen_an}), we found the general applicability of the mechanism where the feedback resurrects oscillations from the AD state. In the case of OD, a proper/suitable form of linear feedback would be helpful to resurrect oscillations or one can also explore the role of nonlinear feedback in such nontrivial OD states. From the results obtained for different cases, we find that the trivial AD state is found to be destabilized through Hopf bifurcation whereas the nontrivial OD states are found to be destabilized even through saddle node type bifurcation (see Appendix \ref{more}). 
\par  In addition to the adaptability of the technique in practical situations, we have illustrated here one more important feature of the technique namely that it does not put any restriction over the number of oscillators contributing towards feedback.  Even with the feedback from a few number of oscillators we can break the death state of the system and thus provide an attractive methodology in practical situations.  Considering a two population network, the contribution from the high frequency oscillators are found to be more preferable compared to the feedback from the low frequency oscillators. 
 
\section*{Acknowledgement}
\par The work of VKC forms part of a research project sponsored by INSA Young Scientist Project.  The work forms part of an IRHPA project of ML, sponsored by the Department of Science Technology (DST), Government of India, who is also supported by a DAE Raja Ramanna Fellowship.  SK thanks the Department of Science and Technology (DST), Government of India, for providing a INSPIRE Fellowship.
\appendix
\section{{\label{gen_an}}Destabilization of AD in a general model}
In this appendix, we show the applicability of the feedback technique over the AD state of a general two coupled system.  For this purpose, we assume $N=2$ and the general forms for ${\bf f}_i$, ${\bf H}$ and ${\bf L}_{ij}$ in Eq. (\ref{gen}), where ${\bf f}_i$ can be chosen as a polynomial in ${\bf{w}_i}$, 
\begin{small}
\begin{eqnarray}
{\bf f}_i({\bf{w}}_i)=\left(\begin{array}{cc}
  F_i(x_i,y_i)+a_i x_i + b_i y_i \\
  G_i(x_i,y_i)+c_i x_i + d_i y_i \\
\end{array}\right), \;
{\bf{w}}_i=\left(\begin{array}{cc}
  x_i \\
  y_i \\
\end{array}\right), \; i=1,2. \quad
\label{fgen}
\end{eqnarray}
\end{small}
In the above, $F_i$ and $G_i$ are nonlinear functions in $x_i$ and $y_i$ and the constants $a_i$, $b_i$, $c_i$ and $d_i$ are system parameters.  The function ${\bf H}({\bf w}_j)$ can be written as
\begin{eqnarray}
 {\bf H}({\bf w}_j)=\bf{w}_j+{\bf \widetilde{H}}({\bf w}_j),
\label{hgen}
\end{eqnarray}
where ${\bf \widetilde{H}}$ is a nonlinear function in ${\bf w}_j$ that introduces nonlinear coupling in the system.  The coupling matrix ${\bf L}_{ij}$ can be taken as
 \begin{small}
\begin{eqnarray}
{\bf L}_{ij}=\left(\begin{array}{cc}
   k_1-\tilde{k}_2 & k_2  \\
 \beta k_2 & \alpha k_1-\beta \tilde{k}_2\\
\end{array}\right),
\label{lgen}
\end{eqnarray}
\end{small}
where $k_1=(-2 \delta_{ij}+1)$, $\tilde{k}_2=\frac{\tilde{\epsilon}}{\epsilon}\delta_{ij}$ and $k_2=\frac{\tilde{\epsilon}}{\epsilon} (-\delta_{ij}+1)$.  The systems are coupled through both direct and conjugate variables, where $k_1$ introduces direct coupling, $k_2$ and $\tilde{k}_2$ introduce conjugate coupling. $\tilde{\epsilon}$, $\alpha$ and $\beta$ are coupling strengths.  The system has a trivial equilibrium point at $(0,0,0,0)$, which may become stable due to the parametric mismatch in the system or due to the coupling in the system.  For example, in \cite{r14} the conjugate coupling in the system induces AD even when the oscillators are identical.  The feedback can be given as ${{\bf u}(t)}= \sum_{k=1}^{N} {  {\bf w}_k}$, where one can also add nonlinear terms in the feedback, if needed:
\begin{small}
\begin{eqnarray}
{\bf g}({\bf u}(t))=\frac{{\bf I}}{2}{{\bf u}(t)},\;\;
\label{ugen}
\end{eqnarray}
\end{small}
\par First, considering the case of coupled identical oscillators $a_i=a$, $b_i=b$, $c_i=c$ and $d_i=d$ ($i=1,2$), the linearization around the trivial equilibrium point $(0,0,0,0)$ can be done.   The eigenvalues corresponding to the case can be obtained easily (note that the nonlinear terms in (\ref{fgen}) and (\ref{hgen}) do not play any role in the linearized equation for the trivial equilibrium point).  The obtained eigenvalues corresponding to the case are of the form
\begin{eqnarray}
\mu_{1,2}&=&\widetilde{\mu}_{1,2}+\eta \label{12}  \\ 
\mu_{3,4}&=&\widetilde{\mu}_{3,4}. 
\label{lamgen}
\end{eqnarray}
In the above $\widetilde{\mu}_{1,2}$ and $\widetilde{\mu}_{3,4}$ are eigenvalues corresponding to the trivial equilibrium point of the system when the feedback is absent ($\eta=0$).  They are given by 
\begin{small}
\begin{eqnarray}
\widetilde{\mu}_{1,2}&=&\frac{1}{2}(a+d-(1+\beta){\tilde{\epsilon}}) \nonumber \\
&&\pm\frac{1}{2} \sqrt{(a-d+(\beta-1)\tilde{\epsilon})^2+4(\beta \tilde{\epsilon}+c)(b+\tilde{\epsilon})}, \quad \\
\widetilde{\mu}_{3,4}&=&\frac{1}{2}(a+d-2(1+\alpha)\epsilon-(1+\beta)\tilde{\epsilon}) \nonumber \\ 
&& \pm \sqrt{(a-d+2(\alpha-1)\epsilon+(\beta-1)\tilde{\epsilon})^2+4\widetilde{C}}, \quad \label{n34} \\
\mathrm{where}\nonumber  \\
\widetilde{C}&=&(c-\beta \tilde{\epsilon})(b-\tilde{\epsilon}) 
\label{mutil}
\end{eqnarray}
\end{small}
\par Depending on the values of the system and coupling parameters, the real part of the eigenvalues ${{\mu_{1,2}}}$ and ${{\mu_{3,4}}}$ in (\ref{12}) and (\ref{lamgen}) are positive or negative while $\eta=0$.  When all the eigenvalues in (\ref{12}) and (\ref{lamgen}) have negative real parts, the stabilization of the equilibrium point gives rise to AD in the system.  From Eq. (\ref{12}), we notice that the increase in $\eta$ causes the eigenvalues $\mu_{1,2}$ to be more positive. Thus a destabilization of the equilibrium point $(0,0,0,0)$ occurs or it wipes off AD.  If the equilibrium point was unstable while $\eta=0$, the increase in $\eta$ never stabilizes the equilibrium point.  Thus the above analysis makes clear the role of $\eta$ in destabilizing the attractor at $(0,0,0,0)$.
\par We can observe a similar effect even in the case where parametric mismatch is present (the case where $a_1 \neq a_2$, $b_1 \neq b_2$, $c_1 \neq c_2$ and $d_1 \neq d_2$) in the system.  The eigenvalues of the system can be obtained by solving the equation
\begin{eqnarray}
\mu^4+A_3 \mu^3+A_2 \mu^2+A_1 \mu+A_0=0, 
\label{eigll}
\end{eqnarray}
where  $A_3=-(a_1+a_2)-(d_1+d_2)+2 \epsilon (1+\alpha)+2 \tilde{\epsilon} (1+\beta)-2 \eta$.  As $A_2$, $A_1$, $A_0$ are not simple in their form, we do not present them here.  Although the eigenvalues $\mu$ obtained from (\ref{eigll}) are not of the simple form, the stability of the equilibrium point $(0,0,0,0)$ in the different parametric regions can be found through the Routh-Hurwitz (R-H) criteria. From the R-H criteria, an equilibrium point is said to be stable only when all the conditions given below are satisfied by the coefficients in the eigenvalue equation (\ref{eigll}).  The R-H criteria are given as
\begin{eqnarray}
&&A_i > 0, \;\; i =0,1,2,3, \nonumber \\
&&A_3 A_2- A_1>0, \nonumber \\
&&A_3 A_2 A_1- A_1^2-A_3^2 A_0>0.
\label{acon} 
\end{eqnarray}
If the coefficients in the characteristic eigenvalue equation (\ref{eigll}) fail to satisfy any one of the condition given above, the equilibrium point becomes unstable.  In this aspect, we consider one of the simpler condition in (\ref{acon}), namely $A_3>0$.   The condition $A_3>0$ is broken when $\eta>\frac{1}{2}\left(-(a_1+a_2)-(d_1+d_2)+2 \epsilon (1+\alpha)+2 \tilde{\epsilon} (1+\beta)\right)$, thus this clearly shows that an increase in $\eta$ destabilizes the equilibrium point $(0,0,0,0)$.  Further, more clear analytical illustration on the role of $\eta$ in the parameter mismatched system is given in Sec. \ref{par_ann2} and \ref{par_ann100} with Stuart-Landau model as an example.
\par As the above type of proof for the non-trivial OD state is too cumbersome, we have illustrated the role of feedback over the state with more examples in the body of the paper as well in Appendix \ref{more} both numerically and analytically (in some cases).  From the above illustrations, one can also notice that in the case of AD, the nonlinear feedback terms cannot play any role (as they lose their significance in the linearized limit) and do not provide any control over it, whereas in the case of OD, the nonlinear feedback also can provide a control over it. 
\section{{\label{more}}Additional Examples}
\subsection{\label{rep}Repulsive link}
\par We consider the case of two Stuart-Landau oscillators coupled diffusively with a repulsive link ($N=2$), as studied in \cite{r28}.   The functions characterizing this equation have the forms,
\begin{figure}[htb!]
\begin{center}
\hspace{-1.0cm}
   \includegraphics[width=9.4cm]{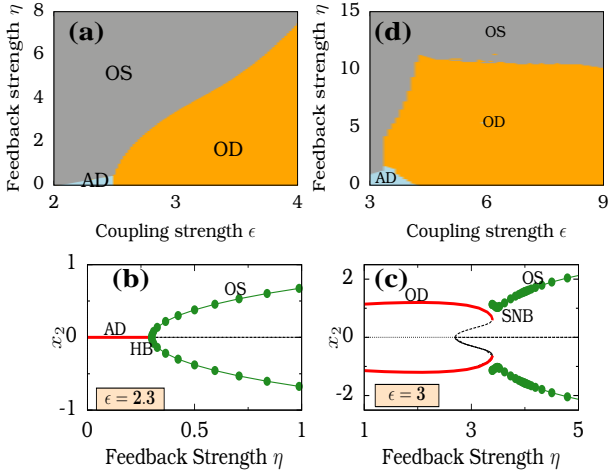}
\end{center}
   \caption{(Color online) (a) Reduction in OD and AD regions of the system (\ref{sys3}) with $\eta$ for $\omega=3.0$.  (b) and (c) show the corresponding transition routes from AD and OD states to OS state.  (d) The emergence of OS regions in the system (\ref{sys4}) with the increase of $\eta$ for $N=200$, $\omega=3.0$ and $p=120$ in (\ref{sys4}). }
\label{re} 
\end{figure}
\begin{small}
\begin{eqnarray}
&{\bf f}_i({\bf{w}}_i)=\left(\begin{array}{cc}
  x_i-\omega y_i-r_i^2x_i  \\
  y_i+\omega x_i-r_i^2 y_i  \\
\end{array}\right),& \quad {\bf H}({\bf w}_j)=\bf{w}_j,  \nonumber \\
&{\bf L}_{ij}=\left(\begin{array}{cc}
   (1-2\delta_{ij}) \delta_{i1}&0  \\
 0&-\delta_{i2}\\
\end{array}\right),&
{\bf g}({\bf u}(t))=\frac{{\bf I}}{2}{{\bf u}(t)},\;\;
\label{sys3}
\end{eqnarray}
\end{small}

The eigenvalues corresponding to the trivial equilibrium point $(0,0,0,0)$ of the system are
\begin{small}
\begin{eqnarray}
\mu&=&\frac{1}{2}(2-\epsilon+\eta) \nonumber \\
&\pm& \sqrt{(\epsilon^2+\eta^2-4\omega^2)\pm 2 \sqrt{\epsilon^2\eta^2+4(\epsilon^2-\eta^2)\omega^2}}.
\end{eqnarray}
\end{small}
We demonstrate the destabilization of this AD state as well as the OD state corresponding to the system (\ref{sys3}) with respect to the feedback in Fig. \ref{re}(a), where we can find that an increase in $\eta$ causes the reduction in AD and OD regions of the system.  The transition route followed by the system, as it transits to oscillatory state, is shown in Figs. \ref{re}(b) and \ref{re}(c).  These figures show that the oscillations are resurrecting from the AD state via Hopf bifurcation as in the previous cases, whereas the resurrection of oscillations from OD state occurs through saddle node bifurcation.  
\begin{figure}[htb!]
\begin{center}
\hspace{-1.0cm}
   \includegraphics[width=9.5cm]{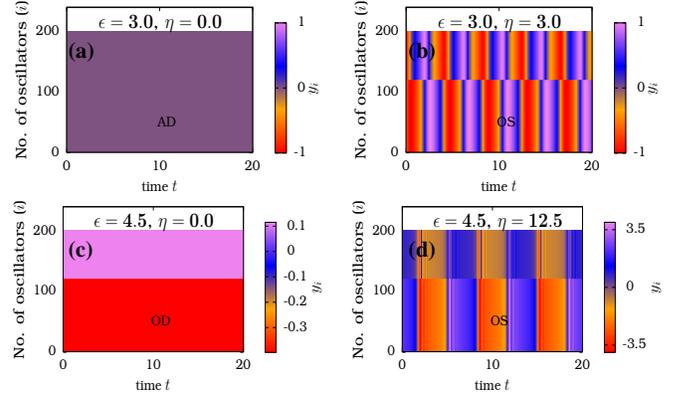}
\end{center}
   \caption{(Color online) Temporal behavior of the system (\ref{sys3}) in the (a) AD state for ($\epsilon, \eta$) $=$ ($3.0,0.0$), (b) the revoked OS state  for ($\epsilon, \eta$) $=$ ($3.0,3.0$), (c) OD state for ($\epsilon, \eta$) $=$ ($4.5,0.0$), and (d) the revoked OS state for ($\epsilon, \eta$) $=$ ($4.5,12.5$).  }
\label{re2} 
\end{figure}

\par Recently, AD and OD in the same system with large number of oscillators ($N$) coupled globally has been seen in \cite{r29}, whose equation is defined by
\begin{small}
\begin{eqnarray}
&&{\bf H}({\bf w}_j)={\bf w}_j, \quad {\bf g}({\bf u}(t))=\frac{{\bf I}}{N}{\bf u}(t)  \nonumber \\
&& {\bf L}_{ij}=\left(\begin{array}{cccc}
   \frac{1}{N}{(1-N \delta_{ij})}&0  \\
 0&-\frac{1}{2}(\sum_{m=1}^{p} \delta_{im}(\delta_{ij}+\delta_{jN}))\\
\end{array}\right)\;\;
\label{sys4}
\end{eqnarray}
\end{small}

Even with $N=200$ oscillators in system (\ref{sys4}), we have shown the reduction in AD and OD regions of the system in Fig. \ref{re}(d). The temporal behavior of the system for different values of $\epsilon$ and $\eta$ are shown in Figs. \ref{re2}(a)-(d).
\subsection{Conjugate coupling}
Next, we consider the case of $N=2$ oscillators coupled through a conjugate coupling described by
\begin{small}
\begin{eqnarray}
{\bf H}({\bf{w}}_j)={\bf{w}}_j, \; 
{\bf L}_{ij}=\left(\begin{array}{cc}
   0&0  \\
{1- \delta_{ij}}&0\\
\end{array}\right), \; 
{\bf g}({\bf u}(t))=\frac{{\bf I}}{2}{\bf u}(t).\;\,
\label{sys5}
\end{eqnarray}
\end{small}

\begin{figure}
\begin{center}
\hspace{-0.2cm}
   \includegraphics[width=9.5cm]{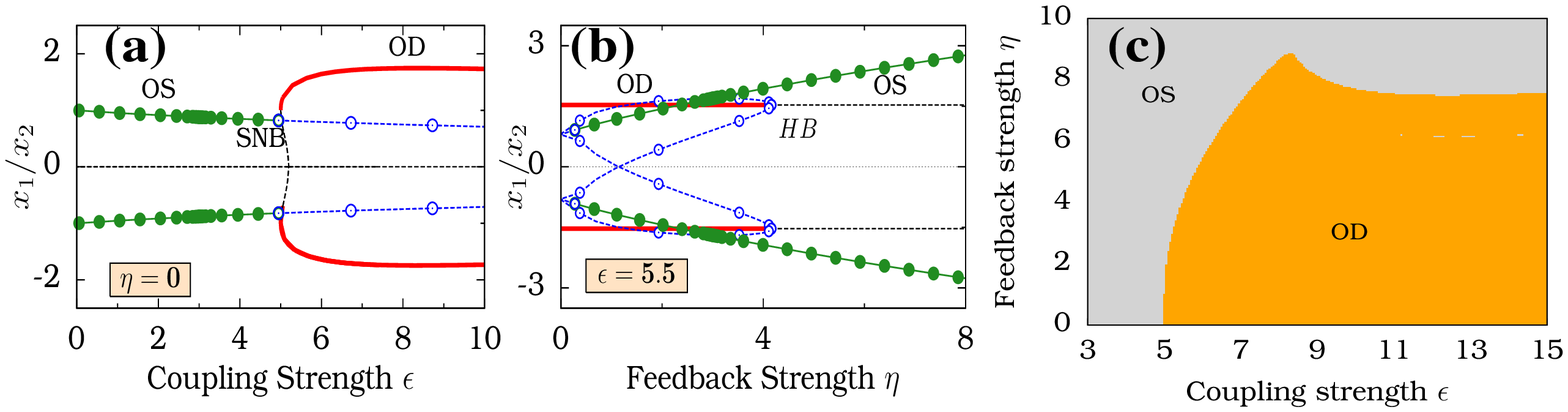}

\end{center}
   \caption{(Color online) (a) The quenching of oscillations through saddle-node bifurcation (SNB) while $\eta=0$ in the case of conjugate coupling.  (b) The destabilization of the OD state with the introduction of feedback via Hopf bifurcation ($HB$). (c) The reduction in the OD regions of the system in the ($\epsilon, \eta$) space for $\omega=5.0$.}
\label{con} 
\end{figure}

This system has a trivial equilibrium point at $e_0$: $(0,0,0,0)$ and has pairs of non-trivial equilibrium points for $\epsilon>\omega$, $e_{1,2}$: $(a_1^* ,b_1^*,-a_1^*,-b_1^*$), where $a_1^*= \pm \sqrt{\frac{1-\omega c}{1+c^2}}$, $b_1^*=-c a_1^*$ and  $e_{3,4}$: $(a_2^* ,b_2^*,-a_2^*,-b_2^*$), where $a_2^*= \pm \sqrt{\frac{1-\omega c}{1+c^2}}$, $b_2^*=c a_2^*$ in which $c=\sqrt{\frac{\epsilon - \omega}{\omega}}$.  The trivial equilibrium point $e_0$ has the eigenvalues

\begin{eqnarray}
\mu_{1,2}&=&(1+\eta) \pm i \sqrt{\omega(\omega+\epsilon)}, \nonumber \\
\mu_{3,4}&=&1\pm i \sqrt{\omega (\omega -\epsilon)}.
\label{eig}
\end{eqnarray}
For $\eta=0$, the equilibrium point $e_0$ is unstable for all values of $\epsilon$ and for $\epsilon=\omega$ a saddle node type bifurcation occurs which stabilizes $e_1$ and $e_2$ as shown in Fig. \ref{con}(a).  For $\eta \neq 0$, the eigenvalues corresponding to the equilibrium point $e_0$ are still unstable, and thus the system is still free of AD.  Then, the equilibrium points $e_1$ and $e_2$ are also destabilized through Hopf bifurcations which is demonstrated in Fig. \ref{con}(b). Then, the OD regions of the system in the ($\epsilon, \eta$) space is shown in Fig. \ref{con}(c), which clearly demonstrates the destabilization of OD with the introduction of feedback.
\subsection{\label{rep2}Repulsive link: van der Pol oscillator}
\begin{figure}
\begin{center}
\hspace{-1cm}
   \includegraphics[width=9.4cm]{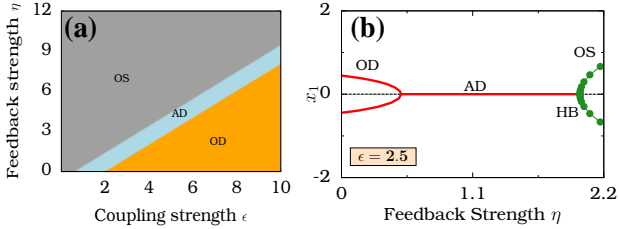}
\end{center}
   \caption{(Color online) (a) shows the AD, OD and OS regions of the system (\ref{sys6}) with respect to $\epsilon$ and $\eta$ for $b=0.5$. (b) The transition route from OD state to OS state via AD state.}
\label{vd} 
\end{figure}

Next, we illustrate the role of feedback in the case two van der Pol oscillators coupled diffusively through a repulsive link. The corresponding dynamical equations are defined through \cite{r15}
\begin{small}
\begin{eqnarray}
&&{\bf f}_i({\bf{w}}_i)=\left(\begin{array}{cc}
 y_i  \\
  b(1-x_i^2)y_i-x_i \\
\end{array}\right),\; \;\;{\bf H}({\bf{w}}_j)={\bf{w}}_j, \nonumber \\
&&{\bf L}_{ij}=\left(\begin{array}{cccc}
   -\delta_{i1} &0  \\
 0&1-2 \delta_{ij} \\
\end{array}\right),\;\,  {\bf g}({\bf u}(t))={\bf Q}{\bf u}(t), \nonumber \\
&&
{\bf Q}=\frac{1}{2}\left(\begin{array}{cccc}
   1 &0  \\
 0&0 \\
\end{array}\right). \quad
\label{sys6}
\end{eqnarray}
\end{small}
The AD and OD regions of the system in the ($\epsilon, \eta$) space are given in Fig. \ref{vd} which show the reduction in the AD and OD regions with respect to $\eta$.  The transition from AD to oscillatory state occurs through a Hopf bifurcation.   On the other hand considering the transition from OD to oscillatory state, the OD state is transformed to AD through inverse pitchfork bifurcation and the oscillatory state arises from the AD state through Hopf bifurcation.  This shows that the role of the feedback in setting oscillations back in the system is not restricted to any particular oscillator. In the next example, we show that the feedback can destabilize oscillation quenching scenario even in chaotic oscillators.
\subsection{Direct and indirect coupling in R\"ossler system}
\begin{figure}
\begin{center}
\hspace{-1cm}
   \includegraphics[width=8.5cm]{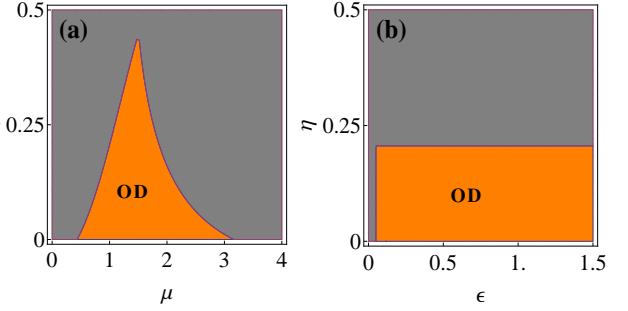}
\end{center}
   \caption{(Color online) Reduction in OD regions of the system (\ref{ros}) (a) in the ($\mu,\eta$) space for $\epsilon=1.0$, (b) in the ($\epsilon,\eta$) space for $\mu=1.0$. In both the figures, we fixed $a=0.1$, $b=0.1$, $c=18.0$ and $k=1.0$.}
\label{rtt} 
\end{figure}

We consider the case of the $N=2$ coupled chaotic R\"ossler system \cite{ros} defined by
\begin{small}
\begin{eqnarray}
&&{\bf f}_i({\bf{w}}_i)=\left(\begin{array}{cc}
 -y_i-z_i  \\
  x_i+a y_i \\
b+z_i(x_i-c)\\
\end{array}\right),\; \;\;{\bf H}({\bf{w}}_j)={\bf w}_j, \nonumber \\
&&{\bf L}_{ij}=\left(\begin{array}{ccc}
  (1-2 \delta_{ij})+\frac{\mu}{N} \frac{v}{x_j} &0&0 \\
 0&0&0\\
0&0&0 \\
\end{array}\right),\;\;  {\bf g}({\bf u}(t))=\frac{{\bf I}}{2}{\bf u}(t). \;\; \nonumber \\
&& \dot{v}=-k v-\frac{\mu}{2} \sum_{j=1}^{N} x_j.
\label{ros}
\end{eqnarray}
\end{small}
Here, the oscillators are coupled directly by diffusive type coupling and are also coupled indirectly to an environment defined by the variable $v$.  This system (\ref{ros}) has only non-zero equilibrium points which include $e_{1,2}$: $(x_i^*,y_i^*,z_i^*,v^*)$ $=$ $(x^*, \frac{-x^*}{a+\eta},\frac{-b}{(x^*+\eta-c)},-\frac{\mu x^*}{k})$, where, $x^*=(-\frac{(\eta-c)}{2} \pm \frac{1}{2}\sqrt{(\eta-c)^2-\frac{4b(a+\eta)}{1+(\eta -\frac{\mu^2}{k})(a+\eta)}})$ .  For the parametric choice $a=0.1$, $b=0.1$, $c=18$ and $k=1$, the equilibrium point $e_2$ is stable.  The OD regions corresponding to the above equilibrium points are given in Figs. \ref{rtt}(a) and \ref{rtt}(b) with respect to the direct coupling strength ($\epsilon$) and with respect to the indirect coupling strength ($\mu$), respectively. The above figures clearly show that the feedback is applicable even for the case of chaotic oscillators.
\subsection{Other chaotic oscillators}
To illustrate further the role of feedback in chaotic oscillators, we consider the $N=2$ coupled Sprott and Lorentz oscillators which are defined respectively by \\

(i)Two coupled Sprott systems with repulsive link \cite{r28},
\begin{small}
\begin{eqnarray}
&&{\bf f}_i({\bf{w}}_i)=\left(\begin{array}{cc}
 -ay_i  \\
  x_i+z_i \\
x_i+y_i^2-z_i\\
\end{array}\right),\; \;\;{\bf H}({\bf{w}}_j)={\bf w}_j, \nonumber \\
&&{\bf L}_{ij}=\left(\begin{array}{ccc}
  (1-2 \delta_{ij})\delta_{j2} &0&0 \\
 0&-\delta_{i1}&0\\
0&0&0 \\
\end{array}\right),\;\;  {\bf g}({\bf u}(t))=\frac{{\bf I}}{2}{\bf u}(t), \qquad
\label{spr}
\end{eqnarray}
\end{small}

\begin{figure}
\begin{center}
\hspace{-1cm}
   \includegraphics[width=9.4cm]{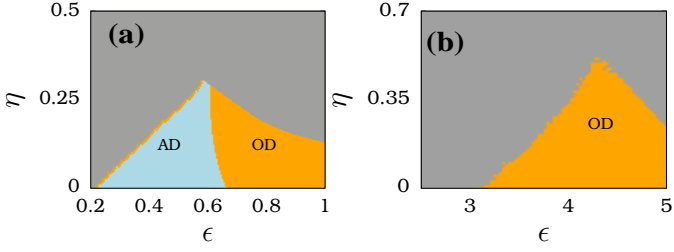}
\end{center}
   \caption{(Color online) Reduction in the AD and OD regions with respect to $\eta$ in the case of (a) the Sprott system (\ref{spr}) for $a=0.225$ and (b) the Lorenz system (\ref{lor}) for $\sigma=10$, $\gamma=28$ and $b=2.67$.}
\label{s} 
\end{figure}
and (ii)Diffusive coupling among two Lorenz oscillators \cite{r2},
\begin{small}
\begin{eqnarray}
&&{\bf f}_i({\bf{w}}_i)=\left(\begin{array}{cc}
 \sigma (y_i-x_i)  \\
  \gamma x_i -y_i -x_i z_i \\
 x_i y_i -b z_i\\
\end{array}\right),\; \;\;{\bf H}({\bf{w}}_j)={\bf w}_j, \nonumber \\
&&{\bf L}_{ij}=\left(\begin{array}{ccc}
  0&0&0 \\
 0&0&0\\
(1-2 \delta_{ij})&0&0 \\
\end{array}\right),\;\;  {\bf g}({\bf u}(t))=\frac{1}{2}{\bf u}(t) . \qquad
\label{lor}
\end{eqnarray}
\end{small}

\par The AD and OD regions corresponding to the coupled Sprott and Lorenz systems ((\ref{spr}) and (\ref{lor})) are given in Fig. \ref{s}, which again confirm that the feedback wipes out AD and OD in the system. 
\section{\label{app2_an}AD region in parametrically mismatched system (\ref{sys1})}
The coefficients in the characteristic eigenvalue equation (\ref{eveq}) are given by 
\begin{small}
\begin{eqnarray}
A_3&=&\epsilon-4-\tilde{a}\eta, \nonumber \\
A_2&=& \frac{1}{4} \tilde{a}^2 \eta^2+\tilde{a} \eta (3 - \epsilon)+(6-3 \epsilon+\omega_1^2+\omega_2^2), \nonumber \\
A_1&=&A_{11} \eta^2+A_{12} \eta +A_{13}, \qquad \nonumber \\
A_0&=&A_{01} \eta^2+ A_{02} \eta+A_{03}, \qquad  \nonumber \\
\mathrm{where}&&\hspace{5cm} \nonumber \\
A_{11}&=&\frac{1}{4} (\epsilon-2) \tilde{a}^2,\nonumber \\
A_{12}&=& (2 \epsilon - 3) \tilde{a}-(\omega_1^2 a_2+\omega_2^2 a_1), \nonumber \\
A_{13}&=& (3 \epsilon-4)+\frac{1}{2} (\epsilon-4)(\omega_1^2+\omega_2^2), \nonumber \\
A_{01}&=&\frac{1}{4} \left[(1-\epsilon) \tilde{a}^2+(\omega_1 a_2+\omega_2 a_1)^2 \right], \nonumber \\
A_{02}&=&\frac{1}{4}\left[(4(1-\epsilon)+\epsilon \omega_1 \omega_2) \tilde{a}-(\epsilon-4)(\omega_1^2 a_2+\omega_2^2 a_1) \right], \nonumber \\
A_{03}&=&(1+\omega_1^2)(1+\omega_2^2)-\frac{\epsilon}{2} (2+\omega_1^2+\omega_2^2).
\label{coeff_ap}
\end{eqnarray}
\end{small}
Here $\tilde{a}=a_1+a_2$.
The eigenvalue equation in (\ref{eveq}) can be solved directly to get the stable regions of the steady state $(0,0,0,0)$.  On the other hand, we can use the R-H criteria to have a closer look at the stable regions of the system (\ref{sys1}). According the R-H criteria, the stable region corresponds to the region in which 
\begin{eqnarray}
&(i)& A_i >0 \qquad i=3,2,1,0 \nonumber\\
&(ii)& A_3 A_2-  A_1>0 \nonumber \\
&(iii)& A_3 A_2 A_1- A_1^2-A_3^2 A_0>0
\label{rhcri}
\end{eqnarray}
\begin{figure}
\begin{center}
   \includegraphics[width=9.5cm]{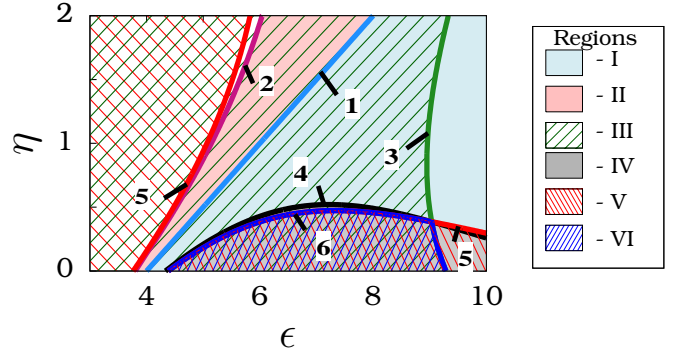}
\end{center}
   \caption{(Color online) The region in which the conditions given in (\ref{rhcri}) are satisfied is depicted in the above figure. The light-blue shaded region (I) which lies under the light-blue curve (curve -$1$) corresponds to $A_3>0$, the region under the pink curve (curve -$2$) which is denoted by region (II) corresponds to $A_1>0$, green shaded region (III) corresponds to $A_0>0$.  The regions satisfying the conditions (ii) and (iii) in (\ref{rhcri}) are the gray shaded region (IV) (the region under curve -$4$) and red shaded region (V) (the region under curve -$5$), separately. The other condition $A_2>0$ is satisfied everywhere in the considered region of ($\epsilon,\eta$).  The region in which all the conditions in (\ref{rhcri}) are satisfied is denoted by blue shaded region (VI) (the region under curve -$6$).}
\label{ap_fig} 
\end{figure}
Now we consider the above criteria one by one and obtain the required stable region of the steady state. \\
(i)a.~~$A_3>0$:~~This condition will be satisfied when 
\begin{eqnarray}
\eta<\frac{\epsilon-4}{\tilde{a}}. 
\label{ib}
\end{eqnarray} 
This simple condition promises that above the value of $\eta=\frac{\epsilon-4}{\tilde{a}}$, the trivial equilibrium point can never be stable and thus oscillation can be revoked by an increase in the value of $\eta$. The region in the $(\epsilon,\eta)$ space in which the above condition is satisfied is denoted by I in Fig. \ref{ap_fig}. \\
(i)b.~~$A_2>0$:~~ When $\epsilon^2-3 \epsilon+3-\omega_1^2-\omega_2^2<0$, if the condition ($A_2>0$) is satisfied for $\eta=0$ then it will be satisfied for all values of $\eta$.  If the condition $A_2>0$ is not satisfied while $\eta=0$ then the condition will not be satisfied for any value of $\eta$.  When $\epsilon^2-3 \epsilon+3-\omega_1^2-\omega_2^2>0$, the condition $A_2>0$ will be satisfied only for the values of $\eta$ given by
\begin{eqnarray}
\eta&>&\frac{2}{\tilde{a}}\left[\epsilon-3 + \sqrt{\epsilon^2-3 \epsilon+3-\omega_1^2-\omega_2^2}\right], \nonumber \\
\eta&<&\frac{2}{\tilde{a}}\left[\epsilon-3 - \sqrt{\epsilon^2-3 \epsilon+3-\omega_1^2-\omega_2^2}\right].
\label{ic}
\end{eqnarray}
This condition $A_2>0$ is satisfied in the whole region considered in Fig. \ref{ap_fig}. \\
(i)c.~~$A_1>0$:~~When $A_{12}^2-4A_{11}A_{13}<0$, if the condition is satisfied for $\eta=0$ then it will continue to be satisfied for all values of $\eta$, if the condition is not satisfied for $\eta=0$, by the variation of $\eta$ also the condition remains to be unsatisfied.  When $A_{12}^2-4A_{11}A_{13}>0$, it will be satisfied only for the values of $\eta$ given by
\begin{eqnarray}
\eta&>&\frac{-A_{12}+\sqrt{A_{12}^2-4 A_{11} A_{13}}}{2 A_{11}}, \nonumber \\
\eta&<&\frac{-A_{12}-\sqrt{A_{12}^2-4 A_{11} A_{13}}}{2 A_{11}}.
\label{id}
\end{eqnarray}
The region of $(\epsilon,\eta)$ in which the above condition is satisfied is denoted by region (II) which lies under pink curve (or curve -$2$) in Fig. \ref{ap_fig}. \\
(i)d.~~$A_0>0$:~~Similar to the previous conditions, if $A_{02}^2-4A_{01}A_{03}>0$, the condition will be satisfied when
\begin{eqnarray}
\eta&>&\frac{-A_{02}+\sqrt{A_{02}^2-4 A_{01} A_{03}}}{2 A_{01}}, \nonumber \\
\eta&<&\frac{-A_{02}-\sqrt{A_{02}^2-4 A_{01} A_{03}}}{2 A_{01}}.
\label{ie}
\end{eqnarray}
Otherwise it will be satisfied for all values of $\eta$ only if the condition is satisfied for $\eta=0$. The region in which the above condition is satisfied is shown in Fig. \ref{ap_fig} as region III.\\
(ii)~~$A_3 A_2- A_1>0$:~~We can find that $A_3 A_2-  A_1$ is a cubic polynomial in $\eta$, and the region in which the above condition is satisfied is given by gray shaded region (IV).  The real roots of $\eta$ satisfying the equation $A_3 A_2-  A_1=0$ forms the boundary of the region. \\
(iii)~~$A_3 A_2 A_1- A_1^2 -A_3^2 A_0>0$:  The region satisfying this condition is shown by red shaded region (V), whose boundary is the solution of the quintic equation $A_3 A_2 A_1- A_1^2-A_3^2 A_0=0$.
\par From Fig. \ref{ap_fig}, we can find that the region satisfying all the criteria given in (\ref{rhcri}) is the region bounded between the curves 
\begin{eqnarray}
&&A_0=0 \;\;\mathrm{or} \;\; \eta=\frac{-A_{02}\pm \sqrt{A_{02}^2-4 A_{01} A_{03}}}{2 A_{01}} \;\; \nonumber \\
\mathrm{and}&&\hspace{0.6cm}\nonumber \\
&&A_3 A_2 A_1- A_1^2-A_3^2 A_0=0.  
\label{bound}
\end{eqnarray}
The region is denoted by blue shaded region VI in Fig. \ref{ap_fig}.  In this region, the equilibrium point $(0,0,0,0)$ is found to be stable or AD occurs in the region.

\end{document}